\documentclass{article}

\newcommand{\f}{\begin{equation}}
\newcommand{\ff}{\end{equation}}

\begin{document}
\renewcommand{\theequation}{\thesection.\arabic{equation}}
\def\sectioneq{{\setcounter{equation}{0}}\section}
\title{%
Triply Special Relativity}
\author{ J.\ Kowalski--Glikman\thanks{{\em Institute for Theoretical
Physics, University of Wroc\l{}aw,  Wroc\l{}aw, Poland}; e-mail
address: {\tt jurekk@ift.uni.wroc.pl}} $\;$ and 
Lee Smolin\thanks{{\em Perimeter Institute for Theoretical Physics, Waterloo, Canada}; e-mail
address: {\tt lsmolin@perimeterinstitute.ca}}  }
 \maketitle
\begin{abstract}
We describe an extension of special relativity characterized by {\it three} invariant scales,
the speed of light, $c$, a mass, $\kappa$ and a length $R$.   This is defined by a non-linear extension
of the Poincare algerbra,  $\cal A$, which we describe here.
For $R\rightarrow \infty$, $\cal A$  becomes the Snyder presentation of the $\kappa$-Poincare algebra, while for
$\kappa \rightarrow \infty$ it becomes the phase space algebra of a particle in deSitter spacetime.
We conjecture that the algebra is relavent for the low energy behavior of quantum gravity,
with $\kappa$ taken to be the Planck mass, for the case of a nonzero cosmogical
constant $\Lambda = R^{-2}$.  We study the modifications of particle motion which follow if
the algebra is taken to define the Poisson structure of the phase space of a relativistic particle.

\end{abstract}

\sectioneq{Introduction}

One of the most fascinating and central questions for contemporary physics is what is the 
symmetry of the low energy limit of quantum gravity.  This question is especially 
interesting once it has been appreciated  that Planck scale effects may be observable. 
This is because present and near future experiments are sensitive to 
 corrections to the basic kinematical relations such as
the energy-momentum relations,  
\f
E^2 = p^2 + m^2 + a l_p E^3 + b l_p^2 E^4 + ...
\label{modified}
\ff
There may also be Planck scale corrections to the conservation laws for  
energy  and momentum  and to the transformation properties of particles under spacetime symmetries. 
Among the possible experimental windows to Planck scale effects are the spectrum of ultra high energy 
cosmic rays, and a possible Planck scale dependence of the speed of light with energy, 
observable in near future observations of gamma ray bursts. 

Neglecting for a moment, the role of the cosmological constant, there are three possibilities.

\begin{itemize}

\item{}{\bf A.}  {\it Poincare invariance}, i.,e,. there is no residue of Planck scale physics in low energy phenomena.

\item{}{\bf B.}  {\it Lorentz symmetry breaking}, so that  
probes sensitive to  Planck scales discover that there really is a preferred reference frame. 

\item{}{\bf C.} {\it Deformed or doubly special relativity (DSR)} \cite{Amelino-Camelia:2000ge}, 
\cite{Amelino-Camelia:2000mn} which refers to the possibility that the 
principle of the relativity of inertial frames may be preserved, but in such a way that
the Planck length or Planck energy becomes an observer independent threshold for
new phenomena.  The name comes from the fact that the symmetry algebra now preserves
two observer independent invariant quantities, the speed of light $c$ and the Planck
length, $l_p$.  

\end{itemize}

An example of this last possibility is $\kappa$-Poincar\'e symmetry \cite{Lukierski:1991pn}, \cite{Lukierski:1991ff}, \cite{kappaM} whose generators satisfy a  non-linear deformation of Poincar\'e invariance, governed by a dimensional parameter
$\kappa = l_p^{-1}$.  This can be understood as the symmetry algebra of a non-commutative
deformation of Minkowski spacetime. Theories invariant under $\kappa-$Poincare
symmetry and other realizations of {\bf DSR} have been constructed and studied in \cite{Kowalski-Glikman:2001gp}, \cite{Bruno:2001mw}, \cite{Magueijo:2001cr}, \cite{Magueijo:2002am}, \cite{Kowalski-Glikman:2002we}.  

These three possibilities are distinguishable experimentally. The second is characterized by 
modified energy momentum relations of the form of (\ref{modified}), but with ordinary
conservation laws of energy and momentum, while possibility {\bf C} is characterized
by non-linear corrections to both energy-momentum relations and conservation laws (see \cite{Amelino-Camelia:2003ex}.)

We have previously conjectured that the third possibility is realized, both  in nature and in the
low energy limit of loop quantum gravity \cite{Amelino-Camelia:2003xp}, \cite{Freidel:2003sp}.   
Support for the second conjecture comes
from $2+1$ gravity coupled to point particles, which is an exactly solvable model. A number
of independent results show that the symmetry algebra which acts on observables of the 
theory is exactly $\kappa-$Poincar\'e symmetry.  One reason to expect that the same
thing will be true of the $3+1$ theory is that modifications of energy-momentum relations
of the form of (\ref{modified}) are seen in several calculations of the propagation of weakly coupled 
excitations of candidates for the ground state of loop quantum gravity. These describe matter
fields or gravitons propagating on flat spacetime, but with modified dispersion relations.
At the same time, it is unlikely that the low energy limit of a quantization of general relativity
can have a preferred frame, as that is ruled out by diffeomorphism invariance, which is instituted by the 
requirement that the states are annihilated by the constraints that generate those 
transformations classically.  

In this note we introduce an extension of doubly special relativity in which the 
Poincar\'e algebra is extended by a third invariant  parameter, which we interpret 
as the cosmological constant, 
$\Lambda$.   Since there are now three observer independent scales, $c$, $l_p$ and
$R=\Lambda^{-1/2}$, we refer to the resulting kinematical theory as {\it triply special relativity.}
In the limit $R \rightarrow 0$, this new algebra reduces to the $\kappa-$Poincar\'e algebra, while
in the limit $l_p \rightarrow 0$ it reduces to the de Sitter (or anti-de Sitter) algebras that
characterize the maximally symmetric solutions with cosmological constant. 

We have both a physical and a theoretical motivation for extending the conjectured 
symmetry algebra of spacetime in this way.   The theoretical motivation begins with the observation
that quantum gravity is unlikely to make sense unless the cosmological constant is a bare
parameter of the theory. This comes from the expectation that $\Lambda$ will be a relevant
parameter that must be controlled to compute the low energy limit of any quantum
theory of gravity. This is certainly true, perturbatively, and there is good evidence it
is true non-perturbatively as well \cite{AL}.  In addition, there is a beautiful argument that connects
the symmetry of the low energy limit of quantum gravity with the symmetry in the presence
of a nonzero cosmological constant \cite{Amelino-Camelia:2003xp}. This arises because it is known that in $2+1$ and 
$3+1$ dimensions the symmetry algebra is quantum deformed, with 
with $z = \ln (q)$
\begin{eqnarray}
z & \approx & \sqrt{\Lambda}l_p  \ \ \ \ \ \mbox{for} \ \ d=2+1 
\ \ \ \cite{witten2+1}, \cite{regge}, \cite{NR1}
\label{2+1}
\\
z & \approx & \Lambda l_p^2  \ \ \ \ \ \mbox{for} \ \ d=3+1  \ \ 
\cite{linking}, \cite{baez-deform}, \cite{qdef}
\label{3+1}
\end{eqnarray}

In the case of $2+1$ gravity, the result that the symmetry algebra is quantum
deformed  when the cosmological constant is turned on 
is rigorous, a complete argument is given in \cite{NR1}.  For the case
of $3+1$ there is good evidence that the local gauge symmetry of the spacetime
connection is quantum deformed from $SU(2)$ to $SU_q(2)$ \cite{linking,baez-deform,qdef}.
In \cite{artem} an argument is given that this extends to the quantum
deformation of the algebra of observables on the boundary of a spacetime with
cosmological constant, so that the subgroup of the de Sitter algebra that
generates the symmetries of the boundary is quantum deformed. This prompts
the conjecture that the algebra of generators that preserve the ground state
of $3+1$ quantum gravity with nonzero $\Lambda$ is quantum deformed. 

We now consider taking the contraction of the quantum deformed symmetry algebra.
The cosmological constant occurs both in the scaling of the translation generators and  in either
(\ref{2+1}) or (\ref{3+1}).  As a result,   the limit
$\Lambda \rightarrow 0$ may be no longer the Poincar\'e algebra.  In the case
of $2+1$ gravity it is exactly the $\kappa-$Poincar\'e algebra \cite{Freidel:2003sp}.   Indeed
this is exactly how the $\kappa$-Poincar\'e algebra was 
found in the first place \cite{Lukierski:1991pn}, \cite{Lukierski:1991ff}.  

In the case of $3+1$ dimensions, one must take into account an additional renormalization
of the energy and momentum generators. This is necessary because, unlike the
case of $2+1$ dimensions, there are local degrees of freedom, and these will
induce a renormalization between the fundamental operators of the theory and the
symmetry generators of the low energy limit. This will be proportional
to a power of the ratio of the ultraviolet and infrared regulator.  Since $LQG$ is known
to be ultraviolet finite, the former is the Planck scale. The latter is of course the cosmological
constant itself.  Thus we have,  

\f
P_{a ,ren} = \left ( {1 \over \sqrt{\Lambda} l_p} \right )^r
\sqrt{\Lambda} M_{5a}
\label{renormalizebis}
\ff

It turns out that for $r<1$ the contraction is the ordinary Poincar\'e algebra, while
for $r=1$ it is again $\kappa-$Poincar\'e.  (For $r>1$ the contraction does not
exist.)  This is supported as well by an explicit calculation\cite{}.  

The physical motivation stems from the observation that there appears to be a
vacuum energy, which can be characterized, so far as all observations done to date, by
a positive cosmological constant, whose value in Planck units is about
\f
\lambda = G\Lambda \hbar \approx 10^{-120}.
\ff

It has, however, proved so far impossible to understand, from known physics, the value
of the observed cosmological constant.   This has remained true despite many attempts.  One may then try 
a new approach to the problem of the cosmological constant by conjecturing that   
$R\approx 10^{60} l_p$ constitutes
a new scale in physics, at which novel, presently unknown laws and principles come into effect. 
But if $R$ is really a scale of new physics,  then we would expect to see surprising phenomena 
in other cases in which the scale appears.  Indeed, there are several such cases, including,

\begin{enumerate}

\item{}The success of the MOND formula, as a phenomenology of galaxy rotation
curves. The situation may be summarized\cite{MOND}  by the statement that the 
need for either dark matter or a modification of Newtonian
dynamics appears whenever the acceleration of a star falls below a critical
value of the acceleration, $a_0$ given roughly by
\f
a_0  = 1.2 \times 10^{-8} \frac{cm}{sec^2} \approx c^2 \sqrt{\Lambda} 
\ff
Whether this indicates the need for a departure from standard physics, or instead, 
just is a phenomenological description of the effects of dark matter is clearly a 
pressing question, but in any case the phenomenology shows that the new 
phenomena is characterized by the scale
of $\Lambda$.  We many note that this means that the scale $\Lambda$ can be
read directly off the date of galactic observations, and in more than one way. It can
be read directly off the data for the Tully-Fisher relation, where it characterizes an
observed relationship between luminous matter and the asymptotic velocity. It can
also be read off of the discrepancy between observed accelerations of stars in galaxies
and those predicted by Newtonian physics based on visible matter. 

\item{}The Pioneer anomaly\cite{Pioneer} consists of the observation of an additional, unexplained
acceleration of all three satellites that have gone outside the solar system towards the
sun, of a magnitude, $ \approx 6 \times a_0$.  

\item{}There is a possible anomalie in $CMB$ observations that can be interpreted
as indicating that the fluctuations of modes with wavelengths greater than $R$ are
suppressed relative to the predictions of the Harrison-Zeldovich spectrum\cite{CMBlow}.

\item{}The possible observations of a time varying $\alpha$ seen by \cite{varyingc}
in quasar absorption line spectra can be interpreted as due to a variation of
the speed of light of order, $\dot{c} \approx  10^{-1} a_0$.  

\end{enumerate}

It is perhaps fair to say that in every case in which we have observational evidence
of phenomena characterized by the scale $R$ there appears to be a divergence from
theoretical expectations.  Of course, some or all of these effects may turn out to be
spurious or have simpler explanations. Still we may take these as hints suggesting we
should look for modifications of physical principles at scales longer than $R$.

In the next section we take an algebraic approach by presenting an extension of the Poincare
algebra characterized by three invariant scales, which we may take to be
$c$, $\kappa = m_p$ and $R= \Lambda^{-1/2}$.  In section 3 we postulate that the algebra 
is the Poisson algebra for a relativistic particle. We study the resulting corrections to the equations
of motion, particularly for the case of circular motion in a central potential. We find violations of the
equivalence principle, and a new force that falls off as $1/distance$, as is the case for MOND.
However, the new force is much too strong for the case of a star in orbit around a galaxy, because non-linear
effects coming from the fact that stars are very large in Planck units overwhelms the naive Newtonian limit. 

\sectioneq{The algebra}

Let us begin with the Poincar\'e algebra. It has as a subalgebra, the Lorentz algebra,
\begin{equation}\label{lorentz}
   [M_{\mu\nu}, M_{\rho\sigma}] = g_{\mu\sigma}\, M_{\nu\rho} + g_{\nu\rho}\, M_{\mu\sigma} - g_{\mu\rho}\, M_{\nu\sigma} - g_{\nu\sigma}\, M_{\mu\rho}
\end{equation}
together with the translations $P_\mu$, which satisfy 
\f
 [ P_\mu , P_\nu ] =0, 
\label{Poincare1}
\ff
to which we add the  action of Lorentz transformations
on  translations
\begin{equation}\label{mp}
    [M_{\mu\nu}, P_{\rho}]= -g_{\mu\rho}\, P_{\nu} + g_{\nu\rho}\, P_{\mu}\quad 
\end{equation}.

This is easily extended to a phase space algebra, by which we mean the combination of the
Poisson algebra for a free relativistic particle and the action of the symmetry generators 
acting on the position and momenta. We then take the commutators to indicate
Poisson brackets so we have 
\f
 [X_\mu, P_\nu] = g_{\mu\nu} 
\label{ccr}
\ff
The algebra is completed by the action of the Lorentz transformations on positions. 
\f
[M_{\mu\nu}, X_{\rho}]= -g_{\mu\rho}\, X_{\nu} + g_{\nu\rho}\, X_{\mu}
\label{mx}
\ff

If we now turn on the cosmological constant $\Lambda= R^{-2}$ the algebra is deformed to 
 de Sitter algebra, which means replacing (\ref{Poincare1}) with
\f
[P_\mu, P_\nu] = \frac1{R^2}\, M_{\mu\nu}
\ff
while the other relations remain unchanged.  

It is useful for what we are about to do to observe that the curvature of position space
is manifested by a non-commutativity of the conjugate variables.  
This of course is well known from basic general relativity.  But seeing it from a phase
space point of view can lead one to ask whether one can do the reverse. That is, could
one deform momentum space to a space of constant curvature? And would this be
manifested by  non-commutativity in the position observables? 

Certainly one can do this. The result is an algebra given by the standard properties
of the lorentz transformations, (\ref{lorentz}), (\ref{mp}), (\ref{ccr}), (\ref{mx})  together with
\f
   [X_\mu, X_\nu] = \frac1{\kappa^2} \, M_{\mu\nu}, \quad  [P_\mu, P_\nu]=0
\label{snyder}
\ff
where we take $l_p = \kappa^{-1}$ to be the Planck scale because it is a small scale deformation of
standard physics.  

Indeed, this is one way of writing the commutation relations that define 
$\kappa$-Poincar\'e symmetry and its action on $\kappa$-Minkowski spacetime.  
In this form it was first written down by Snyder \cite{snyder}. 
Later this was shown \cite{Kowalski-Glikman:2002jr}
to be one basis for the $\kappa$-Poincar\'e algebra, which is now called the Snyder
basis.

A confusing point is that the symmetry algebra generated by the 
$M_{\mu \nu}$ and $P_\mu$ appears to be a classical algebra.  But it acts on
a space of non-commutative coordinates which is otherwise flat. This is confusing 
because  the classical Lie algebras are all
symmetry algebras on classical manifolds (with commuting coordinates) of constant
curvature.   The point is that the relations  (\ref{snyder}) define a particular basis of
a non-trivial Hopf algebra. If one writes the remainder of the Hopf algebra relations
one sees that the algebra is not a classical Lie algebra. 

This  indeed corresponds to the curvature of momentum space, as was shown
in detail by one of us in  \cite{Kowalski-Glikman:2002we}, 
\cite{Kowalski-Glikman:2002jr}, \cite{Kowalski-Glikman:2003we}.  
It should also be mentioned that in the context of quantum groups, the duality between
non-commutativity of the coordinates of the representation space and curvature in the
space of generators was emphasized in the early work of Majid \cite{majidbook}.  

We can now ask if it is possible to do the trick twice. That is, can one make both the 
position and momentum spaces  non-commutative?  One wants then to realize
the standard Lorentz transformation properties (\ref{lorentz}), (\ref{mp}), (\ref{mx})  
and at the same time both
\begin{equation}\label{a3}
    [X_\mu, X_\nu] = \frac1{\kappa^2}\, M_{\mu\nu}, \quad [P_\mu, P_\nu] = \frac1{R^2}\, M_{\mu\nu}.
\end{equation}

This can be done, but it requires deforming also the canonical commutation relation
(\ref{ccr}). One finds by explicit computation that the Jacobi identities are satisfied if one
takes instead 
\begin{equation}\label{tsr}
    [X_\mu, P_\nu] = g_{\mu\nu} - \frac1{\kappa^2}\, P_\mu P_\nu -  
\frac1{R^2}\, X_\mu X_\nu + \frac1{\kappa R}\,\left( X_\mu P_\nu + 
 P_\mu X_\nu + M_{\mu\nu}\right).
\end{equation}

Note that the subalgebras spanned by the pairs $(M, X)$ and $(M, P)$ are just the 
standard de Sitter algebras. Thus we can imagine the phase space as being composed of
the product of  two de Sitter spaces, with in addition a deformed Poisson bracket.  
Alternatively, the entire phase space is now a non-commutative
space.   We see that to do this  gives us an algebra with three universal constants, 
$c$, $\kappa$ and $R$.  

It is also helpful to write the algebra we have found in terms of dimensionless variables
\f
\tilde{X}^\mu = \kappa\, X^\mu , \ \ \ \ 
\tilde{P}_\mu = {R}\,  P_\mu , \\\\\
\ff
The algebra involves only  the dimensionless ratio
\f
r=R\, \kappa
\ff
Again, the standard Lorentz transformation properties are unchanged, while we now have
\f\label{a3}
    [\tilde{X}_\mu, \tilde{X}_\nu] =  \, M_{\mu\nu}, \quad [\tilde{P}_\mu, \tilde{P}_\nu ] =  M_{\mu\nu}.
\ff
\begin{equation}\label{tildetsr}
    [\tilde{X}_\mu,\tilde{P}_\nu] =  r g_{\mu\nu} + M_{\mu\nu} - \frac{1}{r} \left (
\tilde{P}_\mu \tilde{P}_\nu + \tilde{X}_\mu \tilde{X}_\nu - \tilde{X}_\mu \tilde{P}_\nu -  
\tilde{P}_\mu \tilde{X}_\nu \right).
\end{equation}

The algebra $\cal A$ then is given by (\ref{a3}, \ref{tildetsr}) together with the standard
(\ref{lorentz},\ref{mx},\ref{mp}).  In the next section we will be considering it as defining
the Poisson structure on the phase space of a relativistic particle. But it is also
well defined as an operator algebra, with the orderings indicated. 
By extending slightly the construction of Snyder \cite{snyder}, one can find 
representation of $\cal A$ in terms of operators acting on six-dimensional 
Minkowski space with coordinates $\eta_A = (\eta_\mu, \eta_4, \eta_5) = 
(\eta_0, \ldots, \eta_3, \eta_4, \eta_5)$ and the metric of signature  $(+, -, -, \ldots, -)$): 
\begin{equation}\label{a5}
X^\mu = \frac1\kappa\, \left(\eta_4\, \frac{\partial}{\partial \eta_\mu} - 
\eta^\mu\, \frac{\partial}{\partial \eta_4}\right) + \frac{R}{2}\, \frac{\eta^\mu}{\eta_5}
\end{equation}
\begin{equation}\label{a6}
P_\mu = -\frac1R\, \left(\eta_5\, \frac{\partial}{\partial \eta^\mu} - \eta_\mu\, 
\frac{\partial}{\partial \eta_5}\right) + \frac{\kappa}{2}\, \frac{\eta_\mu}{\eta_4}
\end{equation}

\sectioneq{The motion of particles}

Since the formalism we have developed involves the phase space, we can use it
to describe the dynamics of particles.   We then take $\cal A$ as the definition of the Poisson
brackets acting on the phase space $\Gamma = \{ \tilde{X}^\mu, \tilde{P}_\nu \}$.
Our goal in this section is to understand
the physical meaning of  the modifications coming from the deformations of the phase space
algebra parameterized by $l_p$ and $R$. 

The dynamics on the phase space is specified by a reparametrization 
invariant action principle, with the hamiltonian 

\f
H =  N {\cal H}
\ff
where $N$ is the lapse and $\cal H$ is the Hamiltonian constraint.  
The equation of motion for the 
lapse $N$ yields the Hamiltonian constraint,
\f
{\cal H} =0
\label{hamconstraint}
\ff
The equations of motion for positions and momenta are given by
\f
\dot{\tilde{X}}^\mu = N [\tilde{X}^\mu , {\cal H}] , \ \ \ \ 
\dot{\tilde{P}}_\mu = N [\tilde{P} _\mu , {\cal H}]
\ff
subject to the initial data constraint 
(\ref{hamconstraint}). 

For the free particle, the Hamiltonian constraint is given by 
the Casimir of the momentum sector of the phase space algebra, i.e., the 
Casimir of the $(M, P)$ subalgebra. This Casimir reads
\f
 {\cal H}_0 =\tilde{P}_\mu  \tilde{P}^\mu  - M_{\mu \nu}M^{\mu \nu}
\label{freeham}
\ff
It  is easy to see that energy, momentum and angular momentum of particles are
conserved, because,  
\f
[\tilde{P}_\mu, {\cal H}_0] = [M_{\mu\nu}, {\cal H}_0] =0.
\ff

It is also not difficult to verify that, apart from a scaling and ordering, 
the standard definition of the 
Lorentz generators is unchanged,
\f
M_{\mu \nu}= -\frac{1}{r} ( \tilde{X}_\mu \tilde{P}_\nu - \tilde{P}_\mu \tilde{X}_\nu ). 
\label{lorentz-def}
\ff
 
We want to study the question of whether the phenomenology of MOND can be recovered
just from the modifications made so far to dynamics. To do this we add a static potential,
of the form ${\cal U}(\tilde{\rho} )$, where, in the rest frame of the source, 
$\tilde{\rho}^2 = \tilde{X}^i \tilde{X}^i$.  Here we have made a $3+1$ split of
spacetime, with $\tilde{X}^\mu = (\tilde{X}^0 , \tilde{X}^i )$, with $i=1,2,3$.  
Thus, the hamiltonian constraint is now
\f
\label{centralham}
    {\cal H} = \tilde{P}_\mu  \tilde{P}^\mu  - M_{\mu \nu}M^{\mu \nu}+ {\cal U}
\ff
We now compute the equations of motion, using the Poisson brackets (\ref{tildetsr}), (\ref{centralham}).
Using (\ref{lorentz-def}) to simplify the resulting expressions we find that
\begin{eqnarray}
\dot{\tilde{X}}_\mu &=&  2N \tilde{P}_\mu \left [
r+ \frac{1}{r} ( 2 \tilde{X}\cdot \tilde{P} - \tilde{X}^2 -\tilde{P}^2  ) 
+ \frac{1}{2r} \tilde{X}^\lambda \frac{\partial {\cal U}}{\partial \tilde{X}^\lambda } 
\right ] 
\nonumber \\
&&- \frac{N}{r} \tilde{X}_\mu \tilde{P}^\lambda \frac{\partial {\cal U}}{\partial \tilde{X}^\lambda } 
\\
\dot{\tilde{P}}_\mu &=& - Nr \frac{\partial {\cal U}}{\partial \tilde{X}^\mu  } + \frac{N}{r}
\left [ \tilde{X}_\mu ( 2\tilde{P}^\lambda - \tilde{X}^\lambda    ) -\tilde{P}_\mu \tilde{P}^\lambda
\right ] \frac{\partial {\cal U}}{\partial \tilde{X}^\lambda } 
\end{eqnarray}

We now impose conditions that single out circular motion. These are
\f
\tilde{P}^\lambda \frac{\partial {\cal U}}{\partial \tilde{X}^\lambda } =0 , \ \ \ \ 
\tilde{P} \cdot \tilde{X} = - \tilde{E}\tilde{t}
\ff
We also posit that the potential is Newton's gravitational potential
\f
{\cal U}= m^2 + c\frac{GMm}{\tilde{\rho}}
\ff
where $c$, like $N$ is to be determined by matching to the non-relativistic Newton's
laws and $M$ is the mass of the central body. 

These reduce the equations of motion for the spatial components to
\begin{eqnarray}
\dot{\tilde{X}}_i &=& 2N r \tilde{P}_i \left [
1- \frac{1}{r^2} ( \tilde{X}^2 +\tilde{P}^2 -2 \tilde{E}\tilde{t}  ) 
+ \frac{GcMm}{2r^2 \tilde{\rho} }
\right ] 
\\
\dot{\tilde{P}}_i& =& - \frac{NGcMm \tilde{X}_i}{\tilde{\rho}}   
\left [  \frac{r}{\tilde{\rho}^2} +  \frac{1}{r \tilde{\rho}}  \right ] 
\end{eqnarray}
We now choose $N$ so that for the physical, dimensional variables
\f
m \dot{X}_i = \tilde{P}_i  
\ff
This requires
\f
N=\frac{1}{2mAR^2}
\label{Ndef}
\ff
where
\f
A= 1- \frac{1}{r^2} ( \tilde{X}^2 +\tilde{P}^2 -2 \tilde{E}\tilde{t}  ) 
+ \frac{GcMm}{2r^2 \tilde{\rho}}
\ff
Using (\ref{Ndef}), $\dot{\tilde{P}}_i$ becomes
\f
\dot{\tilde{P}}_i =  - \frac{GcMl_p^2  \tilde{X}_i}{2R^2 A \tilde{\rho}}   
\left [  \frac{r}{\tilde{\rho}^2} +  \frac{1}{r \tilde{\rho}}  \right ] 
\ff
Combining these we find the acceleration is
\f
\ddot{\tilde{X}}_i= -\frac{GcM \tilde{X}_i}{2R^3Al_p \tilde{\rho}}  
\left [  \frac{r}{\tilde{\rho}^2} +  \frac{1}{r \tilde{\rho}}  \right ] 
\ff
We now go back to dimensionless variables.  We note that 
$A= 1 + (...)/R^2$, so that as $R\rightarrow \infty $ with all other variables held fixed, $A \rightarrow 1$. Thus, it is 
natural to expect that $A$ contains corrections which are unimportant except on cosmological scales. 
We therefor choose $c$ so that the Newtonian limit is obtained as $R\rightarrow \infty$, so that
\f
\ddot{X}_i = -\frac{GM \hat{X}_i}{ A }  
\left [  \frac{1}{{\rho}^2} +  \frac{l_p}{R \rho}    \right ] 
\ff
This fixes
\f
c=\frac{2R^2m}{l_p}
\ff
Assuming that $A \approx 1$, we  do find an aparent MOND-like force, which
is the term that falls off like $1/\rho$. However this is much too small, and it only
becomes comparable to the Newtonian force for $\rho \approx R^2 / l_p$.   We also fail to see the emergence of 
a critical acceleration scale.  

However, this is only the case if the masses are very small. For it is easy to see that the effect
of the $A$ term leads to drastic violations of the equivalence principle. Consider the term
in $A$ proportional to 
\f
z=\frac{GcMm}{2r^2 \tilde{\rho}} = \frac{GMm^2 l_p^2}{\rho}= \frac{GM}{\rho} \left(\frac{m}{m_p}\right)^2 
\ff
For a proton, in orbit around a galaxy, $z \approx 10^{-40}$. But the situation for a star around a galaxy is
very different. In this case,   the second factor overwhelms the first, so that $z \approx 10^{72}$
Thus, since $A  \approx 1 +z +... \approx z$, the Newtonian limit is not obtained for the case of 
a star in circular orbit in a galaxy, instead we find an acceleration
\f
\ddot{x}_i = -\frac{\hat{x}_i}{ m^2 l_p^2 \rho  }  
\ff
which is very far from the Newtonian limit.  

The lesson is that due to the non-linearities in the algebra, there are corrections to the dynamics that lead
to massive violations of the equivalence principle. We may fix the constants so that Newton's laws are
satisfied for masses much less than the Planck mass. This happens because the standard terms in
the Poisson brackets dominate. But for stars in orbit around a galaxy, the new terms in the 
Poisson brackets such as the $M_{\mu \nu}$ term in (\ref{tsr}) are much more important than the
conventional $\eta_{\mu\nu}$ terms. The reason is that factors like $(m/m_p )^2 \approx 10^{76}$
for a star can overcome suppressions of order $l_p/R \approx 10^{60}$.  

Related to this is the observation that since the algebra is non-linear, it is no longer true that the description of 
a composite system follows in a simple way from the action on the constituents. It is straightforward to 
show that if $\cal A$ is posited as the Poisson algebra for elementary particles, it will not be satisfied 
for the total momentum and center of mass coordinates of a composite system, if they are given by the usual
linear formulas of standard mechanics.  We expect that this is related to similar issues that arise
in the application of $DSR$ to composite systems.  These questions must be resolved before it can be
determined whether the symmetry algebra described here may or may not be relevant for real physics.

\sectioneq*{Acknowledgement}

We would like to thank Giovanni Amelino-Camelia, Michele Arzano, Laurent Freidel, Fotini Markopoulou, 
Stacy  MacGaugh, Joao Maguiejo, John Moffat, and Artem Starodubtsev for discussions on various
related questions.   JKG would like to thank Perimeter Institute for the hospitality in December 2003, 
when this project started and in June 2004, when it was completed.

\end{document}